\documentstyle[amssymb,preprint,prb,aps]{revtex}
\tightenlines
\begin{document}
\draft
\title{Direct Observation of Coexistence of Ferromagnetism and 
Superconductivity in RuSr$_{2}$(Gd$_{0.7}$Ce$_{0.3}$)$_{2}$Cu$_{2}$O$_{10}$}
\author{S. Y. Chen$^{*,\dagger}$, J. Shulman$^{\dagger,\S}$, 
Y. S. Wang$^{*,\dagger}$, D. H. Cao$^{\dagger}$, C. Wang$^{*,\dagger}$,
Q. Y. Chen$^{*,\dagger}$,\\ T. H. Johansen$^{\dagger,\|}$, 
W. K. Chu$^{*,\dagger}$ and C. W. Chu$^{*,\dagger,\ddagger}$}
\address{$^{*}$Department of Physics, University of Houston, Houston, TX 
77204-5506\\
$^{\dagger}$Texas Center for Superconductivity, University of Houston, 
Houston, TX 77204-5932\\
$^{\ddagger}$Lawrence Berkeley National Laboratory, 1 Cyclotron Road, 
Berkeley, CA 94720\\
$^{\S}$Permanent address: Joint Science Department, W. M. Keck Science Center,
925 N. Mills Ave., Claremont, CA 91711-5916\\
$^{\|}$Permanent address: Department of Physics, University of Oslo,
Box 1048 Blindern, N-0316, Oslo, Norway}
\date{\today}
\maketitle

\begin{abstract}
Recent reports of the detecting of ferromagnetism and superconductivity 
in ruthenium-cuprates have aroused great interest. Unfortunately, whether 
the two antagonistic phenomena coexist in the same space in the compounds 
remains unresolved. By employing the magneto-optical-imaging technique, 
ferromagnetism and superconductivity were indeed directly observed to 
coexist in the same space in 
RuSr$_{2}$(Gd$_{0.7}$Ce$_{0.3}$)$_{2}$Cu$_{2}$O$_{10}$ 
within the experimental resolution of $\sim$ 10 $\mu$m. The observation 
sets a length scale limit for models proposed to account for the 
competition between ferromagnetism and superconductivity, especially 
$d$-wave superconductivity, in this interesting class of compounds.
\end{abstract}





% body of paper here

\section{INTRODUCTION}
The antagonistic nature between ferromagnetism and superconductivity has long 
been recognized.\cite{Bulaevskii} By replacing the CuO-chain layers in the 
charge reservoir blocks of the cuprate high temperature superconductor with 
RuO$_{2}$ layers, the ruthenium-cuprate compounds have been formed. On cooling, 
they have recently been reported \cite{Bauernfeind,Felner1,Chmaissem,Meng} to 
undergo a magnetic transition at a temperature $T_{m}$ followed by a 
superconducting transition at a lower temperature $T_{s}$. For instance, 
the transition temperatures ($T_{m}$, $T_{s}$) are (90-180~K, 30-40~K) and 
(130-133~K, 30-45~K) for Ru-1222 
[RuSr$_{2}$(Gd$_{0.7}$Ce$_{0.3}$)$_{2}$Cu$_{2}$O$_{10}$, 
RuSr$_{2}$(Eu$_{0.7}$Ce$_{0.3}$)$_{2}$Cu$_{2}$O$_{10}$],\cite{Felner1} 
and Ru-1212 [RuSr$_{2}$GdCu$_{2}$O$_{8}$, 
RuSr$_{2}$EuCu$_{2}$O$_{8}$],\cite{Chmaissem,Meng} respectively. The 
appearance of a spontaneous magnetic moment in these compounds below $T_{m}$ 
at a very low field suggests that the transition at $T_{m}$ must have a 
significant ferromagnetic component, in spite of the recent detection of an 
antiferromagnetic order associated with the Ru-sublattice of Ru-1212 by a 
neutron diffraction experiment \cite{Lynn} in a field below 1~T. Ferromagnetism 
and superconductivity have thus been proposed to coexist in Ru-1222 and -1212, 
an extremely unusual occurrence. Although magnetic studies have unambiguously 
shown a weak ferromagnetic order below $T_{m}$, the superconducting transition 
below $T_{s}$ has only been demonstrated by resistivity measurements in 
ruthenium-cuprate polycrystalline samples and without a bulk Meissner 
effect,\cite{Chu1} the usual signature of a superconducting transition. 
Therefore, questions remain as to whether the two phenomena coexist in the 
same location in the samples below $T_{s}$ and, if so, what the structures of 
these states are, especially below $T_{s}$. This is particularly true in view 
of the absence of a bulk Meissner effect below $T_{s}$ \cite{Chu1} and the 
questioning of the very existence of superconductivity in 
Ru-1212.\cite{Felner2} To address the first question, we have carried out 
space-resolved magneto-optical imaging of polycrystalline samples of 
Ru-1222 [RuSr$_{2}$(Gd$_{0.7}$Ce$_{0.3}$)$_{2}$Cu$_{2}$O$_{10}$] between 
5 and 300~K in their superconducting, weak ferromagnetic, and paramagnetic 
states. The results clearly demonstrate that ferromagnetism and 
superconductivity do coexist in the same location in the samples examined 
within our experimental resolution of $\sim$ 10 $\mu$m. The observation sets 
a new length scale limit for models proposed to account for the competition 
between ferromagnetism and superconductivity, especially $d$-wave 
superconductivity, in this interesting class of compounds, namely, 
superconducting ferromagnets, in which $T_{m} > T_{s}$, in contrast to the 
ferromagnetic superconductors, where $T_{s} > T_{m}$, previously investigated.

\section{EXPERIMENTAL}
The Ru-1222 samples studied were prepared by the standard solid-state reaction 
of thoroughly mixed powders of RuO$_{2}$ (99.95\%), SrCO$_{3}$ (99.99\%), 
Gd$_{2}$O$_{3}$ (99.99\%), CeO$_{2}$ (99.99\%), and CuO (99.9\%), with the 
cation ratios of Ru:Sr:(Gd$_{0.7}$Ce$_{0.3}$):Cu = 1:2:2:2. Details of sample 
preparation and its relation to the superconducting and magnetic properties 
of the samples will be published elsewhere.\cite{Cao} The structure was 
determined by powder X-ray diffraction (XRD) using the Rigaku DMAX-IIIB 
diffractometer; the composition by the energy dispersive analysis of X-ray 
(EDAX); the resistivity ($\rho$) by the standard four-lead technique, 
employing the Linear Research Model LR-700 Bridge; the magnetization 
($M$) by the Quantum Design SQUID magnetometer; and the magneto-optical 
imaging (MOI) by a system similar to the one described 
previously.\cite{Johansen} Our MOI system, which uses indicator films of 
Bi-substituted yttrium-iron-garnet (Bi:YIG) with in-plane magnetization, 
consists of an Olympus polarizing microscope, an Olympus Magnafire 
Imaging System, and an Oxford Microstat. For comparison with MOI pictures, 
the surface morphology of the sample was observed at room temperature with a 
scanning electron microscope (SEM).

\section{RESULTS AND DISCUSSION}
The powder XRD pattern in Fig.~\ref{fig1} shows that the Ru-1222 sample is 
rather pure but has slight traces of possible impurities of SrRuO$_{3}$ and 
Gd$_{2}$Ru$_{2}$O$_{7}$. The Ru-1222 phase exhibits a tetragonal structure 
with lattice parameters: $a = 3.841(2)$ and $c = 28.62(1)$, in good agreement 
with previous reports.\cite{Bauernfeind,Felner1} The EDAX data show a uniform 
composition across the samples to a spatial resolution of 1--2~$\mu$m. 
Figure~\ref{fig2} shows the temperature dependence of $\rho$ of Ru-1222 at 
ambient as well as at 5~T. The sample shows a metallic behavior 
above $T_{s}$, a sudden $\rho$-drop with an onset temperature $\sim 
38$~K, and zero-$\rho$ temperature $\sim 28$~K, which is broadened and 
shifted toward a lower temperature by a magnetic field, characteristic of a 
superconducting transition. The low-field magnetic susceptibility 
($\chi$) at 1.2~Oe in both the zero-field-cooled (ZFC) and field-cooled (FC) 
modes is given as a function of temperature in Fig.~\ref{fig3}. A large 
diamagnetic shift is observed in the ZFC-$\chi$ below $T_{s} \sim 30$~K, 
representing a large superconducting shielding in the sample and consistent 
with the $\rho$-results. The behavior of ZFC-$\chi$ at temperatures above 
$T_{s}$ shows a magnetic transition near $T_{m} \sim 90$~K, although the shape 
of ZFC-$\chi$ above $T_{s}$ depends on the detailed nature of the magnetic 
state and the field-history of the measurement. The FC-$\chi$ at low field 
displays a large upturn at $T_{m} \sim 90$~K, proceeded by a small rise 
at $\sim 130$~K, similar to that previously observed.\cite{Bauernfeind} 
The low-field FC-$\chi$ rise at $T_{m} \sim 90$~K shows that a spontaneous 
magnetic moment appears below $T_{m}$, indicative of a weak ferromagnetic 
transition, in agreement with the previous report.\cite{Bauernfeind} A small 
increase at $\sim 130$~K is also evident and may be associated with the 
magnetic impurity phase of SrRuO$_{3}$, Gd$_{2}$Ru$_{2}$O$_{7}$, or other 
reasons to be described later. In contrast to an earlier 
observation,\cite{Felner1} FC-$\chi$ displays a slight drop in our Ru-1222 
samples near $T_{s} \sim 30$~K, similar to a superconducting Meissner 
transition of a small volume fraction, prior to its resumption of a small 
increase below $\sim 22$~K. However, the magnitude of such a diamagnetic shift 
in the FC-$\chi$ was found to depend on the sample and is rapidly suppressed 
by an external field. It becomes zero in fields above $\sim 5$~Oe for the 
sample shown in Fig.~\ref{fig3}, reminiscent of a transition associated with 
the phase-lock of an aggregation of small Josephson-coupled superconducting 
grains or domains. Recently, a similar diamagnetic shift in the 
FC-$\chi$ was also detected in Ru-1212, but was attributed to a possible 
spontaneous-vortex-state to Meissner-state transition on cooling.\cite{Bernhard}

We have also examined the superconducting remnant state of the sample, which 
was achieved by cooling the sample to its superconducting state to 5~K in the 
absence of a magnetic field, followed by increasing the field to 560~Oe 
(for reasons that will be evident later) to reach its critical state, and 
finally reducing the field back to zero. A magnetic field is thus trapped 
by the sample in its remnant state due to the persistent supercurrent at 5~K. 
As the sample is warmed up, the trapped field is expected to decrease to zero 
at $T_{s}$ in accordance with the decrease of critical current in the sample 
with increasing temperature. This was indeed observed as shown in the inset 
to Fig.~\ref{fig3}, except that the residual field vanishes not at $T_{s}$, 
but only above $T_{m}$. This is attributed to the fact that the magnetization 
shown in the inset to Fig.~\ref{fig3} consists of two contributions: the 
persistent supercurrent that vanishes at $T_{s}$ and the ferromagnetic moment 
that vanishes only at $T_{m}$. However, the former decreases with increasing 
temperature to $T_{s}$ at a much greater rate than the latter.

The MOI technique \cite{Johansen} is employed to ``see'' directly the 
magnetism generated by the Ru-1222 sample. The imaging is based on the 
large Faraday effect in the garnet film, which is mounted in direct contact 
with the sample. The optical arrangement is such that the incoming 
plane-polarized light is rotated proportionally to the local magnetic 
field on the sample surface, and by crossing the analyzer an image is 
formed where the brightness directly corresponds to the local value of the 
magnetic field. The spatial resolution of the present system is better than 
10~$\mu$m.

The Ru-1222 samples for MOI were dry-polished with 0.3~$\mu$m sandpaper. 
To monitor the evolution of the magnetic moment we cooled the sample in 
external fields ($H$) of $\sim$~0.5, 14, and 83~Oe and determined the MOI 
images of the Ru-1222 in its paramagnetic, ferromagnetic, and superconducting 
states. The typical results at 83~Oe, clearer than but similar to those at 
lower fields, are shown in Figs.~\ref{fig4}a-c with the relative brightness 
proportional to the magnetic field generated by the sample. In the pictures, 
one should ignore the sharp-edged contrasts, which are domain boundaries 
intrinsic of the Bi:YIG indicator film. At 95~K $> T_{m} \sim 90$~K, where $M$ 
is very small, the magnetic induction of the sample $B = (H + 4\pi M) 
\cong H$, and the sample is thus indistinguishable from its background 
(Fig.~\ref{fig4}a). At 62~K $< T_{m}$, $M$ has a large positive value 
and $B$ becomes much greater than $H$. The sample becomes brighter than the 
background (Fig.~\ref{fig4}b). At 5~K, no decrease of sample brightness was 
detected and, instead, the sample became even brighter (Fig.~\ref{fig4}c). 
This is in agreement with the FC-$\chi$ data (Fig.~\ref{fig3}), where the 
magnetic moment at 5~K is greater than that at 62~K and the small drop in 
moment at $\sim 22$~K vanishes at the measuring field of 14~Oe. It should be 
noted that even at the weak earth field of $\sim 0.5$~Oe, a bright sample 
image was still detected, indicative of the existence of magnetic flux in 
the sample in its superconducting state. This is consistent with the previous 
suggestion \cite{Chu1} of the absence in Ru-1212 of a bulk Meissner state. 
In Figs.~\ref{fig4}b-c, bright, granular magnetic structures are clearly 
observed below $T_{m}$. This is attributable to the granular structure of the 
polycrystalline sample as revealed by our SEM data. There is little difference 
in the images below $T_{m}$, suggesting that these structures are mainly due 
to the magnetic contribution in the sample. There should be a superconducting 
contribution to the magnetic behavior of the sample during field cooling to 
below $T_{s}$. However, the increasing brightness of these granular structures 
with lowering temperatures below $T_{s}$, which indicates a strong magnetic 
field due to ferromagnetism, prevents us from separating the superconducting 
from the magnetic contribution in the sample.

To identify the superconducting behavior of the Ru-1222 sample, we recorded 
MOI images of the same sample in its superconducting remnant state, as 
described earlier, at different temperatures. As pointed out earlier, in the 
remnant state, the field trapped in the sample is associated with the 
persistent supercurrent and is thus expected to generate a bright structure 
corresponding to the superconducting parts of the sample. The superconducting 
remnant state at 5~K was initially obtained by the application and the 
subsequent removal of a field of 576~Oe that is strong enough to generate 
a magnetic granular structure resolvable by our MOI system near $T_{s}$. The 
MOI results of the Ru-1222 sample are shown in Figs.~\ref{fig5}a-c, with the 
relative brightness proportional to the strength of the field trapped. 
Indeed, a bright granular structure was observed at 5~K $< T_{s}$ 
(Fig.~\ref{fig5}a). When the sample is warmed up, the trapped field decreases 
rapidly and continuously due to the decreasing persistent supercurrent, 
as evidenced by the rapidly diminishing brightness of the granular structure 
(Fig.~\ref{fig5}b). At temperatures above $T_{s}$, the granular structure 
disappears completely (Fig.~\ref{fig5}c), showing directly that it is caused 
by superconductivity. This cannot be associated with a remnant magnetism of 
the sample because the brightness decreases too rapidly. This is in agreement 
with our magnetization results of the sample, in its superconducting remnant 
state achieved in a similar field, that is shown in the inset to 
Fig.~\ref{fig3}.

Finally, to determine whether the superconductivity and ferromagnetism 
originate from the same place in the sample, we decided to compare the 
granular structures caused by the ferromagnetism and superconductivity, 
respectively, at a higher magnification with more enhanced brightness and 
contrast. Figure~\ref{fig6}a shows the superconducting granular structure of 
the Ru-1222 sample obtained in its remnant state at 5~K. Figure~\ref{fig6}b 
displays the ferromagnetic granular structure of the same area 
on the sample in its ferromagnetic state at 62~K. Both pictures are 
obtained from the same rectangular areas marked in Figs.~\ref{fig5}a and 
\ref{fig4}b, respectively. It is clear that the two structures are essentially 
identical within the resolution of our MOI system, \textit{i.e.}~they almost 
fall on top of each other. The difference in the brightness of the two 
structures is due to the different magnetic field strengths generated by the 
two states. Therefore, the observation directly demonstrates that 
superconductivity and ferromagnetism do occur in the same location in the 
Ru-1212 sample within a resolution of $\sim$ 10~$\mu$m.

Many studies have been carried out on the nature of this superconducting state 
in the (weak) ferromagnetic 
background.\cite{Chu1,Bernhard,Sonin,Knigavko,Pickett} Depending on the 
relative strengths of the superconducting and ferromagnetic interactions, 
various transition sequences have been proposed between the paramagnetic, 
(weak) ferromagnetic, spontaneous vortex, and Meissner phases in the 
superconducting ferromagnets.\cite{Bernhard,Sonin,Knigavko,Pickett,Chu2} The 
failure to detect a bulk superconducting state nor a superconducting 
condensation energy led to the suggestion \cite{Chu1,Chu2} of a possible novel 
crypto-superconducting state in the superconducting ferromagnet, Ru-1212. Such 
a state can have a fine granular microstructure beset by the ferromagnetic 
walls between the antiferromagnetic ``domains,'' or a non-uniform filamentary 
structure existing in the less magnetic walls between the ferromagnetic 
domains. This appears to be consistent with a recent model 
calculation. \cite{Zhu} However, based on the recent observation of a 
diamagnetic shift in the FC-$\chi$, a paramagnetic $\rightarrow$ (weak) 
ferromagnetic $\rightarrow$ spontaneous-vortex $\rightarrow$ Meissner phase 
transition sequence in Ru-1212 upon cooling has also been 
proposed.\cite{Bernhard} Unfortunately, the magnitude of the diamagnetic shift 
in Ru-1212 near 30~K decreases rapidly with an applied magnetic field and drops 
to zero at $\sim 12$~Oe, similar to that observed here in Ru-1222, reminiscent 
of a phase-lock transition of an aggregate of superconducting fine grains. 
In view of the ubiquitous electronic phase separation \cite{Tranquada,Teresa} 
in the underdoped superconducting cuprates and the colossal magnetoresistant 
manganites, we also envision a possible similar phase separation in these 
underdoped Ru-1212 and -1222 samples near or below their magnetic transition, 
leaving an electronically non-uniform magnetic system. Such a system can have 
nanoscale interdispersions of different ferromagnetic strengths with 
superconductivity residing in the less magnetic (or even antiferromagnetic) 
dispersions. While the present investigation cannot distinguish one scenario 
from the other mentioned above, it sets a limit on the length scale of the 
superconducting grains or domains that is much less than 10~$\mu$m. It should 
be noted that the superconducting grains or domains to which we refer here 
are considered to be part of and thus smaller than the crystalline grains 
revealed by the SEM and/or MOI data. Further refinement in the length scale 
depends critically on the availability of single-crystalline and/or epitaxial 
thin-film samples of Ru-1222 and -1212. By fine-tuning the magnetic and 
superconducting interactions, superconducting ferromagnets will provide a 
unique opportunity for the study of the interplay between magnetism and 
superconductivity, and particularly between ferromagnetism and $d$-wave 
superconductivity in cuprates.

% \section*{Acknowledgments}

\acknowledgments
The work at Houston is supported in part by the National Science Foundation, 
the T.~L.~L. Temple Foundation, the John and Rebecca Moores Endowment, the 
Superconductivity Quasi Endowment Fund, and the State of Texas through the 
Texas Center for Superconductivity at the University of Houston; and at 
Berkeley by the Director, Office of Energy Research, Office of Basic 
Energy Sciences, Division of Materials Sciences, of the U.~S. Department 
of Energy. One of the authors (THJ) is grateful to the Norwegian Research 
Council for financial support.

% now the references. delete or change fake bibitem. delete next three
%   lines and directly read in your .bbl file if you use bibtex.

\begin{figure}[tbp]
\caption{The XRD pattern of Ru-1222 sample: * - impurities.}
\label{fig1}
\end{figure}

\begin{figure}[tbp]
\caption{$\rho$(T) of Ru-1222: (a) 0~T and (b) 5~T.}
\label{fig2}
\end{figure}

\begin{figure}[tbp]
\caption{ZFC - $\chi$(T) (a) and FC - $\chi$(T) (b) at 1.2~Oe: 
Inset - The decay of the superconducting remnant moment achieved at 
5~K after the application and subsequent removal of 560~Oe. The 
minimum $M$ that will give discernable magnetic bulk ($\cdot\cdot\cdot$) 
or granular structure (--- - --- -) by the MOI technique is also given.}
\label{fig3}
\end{figure}

\begin{figure}[tbp]
\caption{MOIs of Ru-1222 field cooled in 83~Oe at (a) 95 K, (b) 62 K, and 
(c) 5 K.}
\label{fig4}
\end{figure}

\begin{figure}[tbp]
\caption{MOIs of Ru-1222 in its remnant state achieved at a maximum field of 
576 Oe was applied on warming in zero field at (a) 5~K, (b) 20~K, and (c) 40~K.}
\label{fig5}
\end{figure}

\begin{figure}[tbp]
\caption{Comparison between the superconducting granular structure (a) 
at 5~K 
with the magnetic granular structure (b) at 62~K in Ru-1222. They are the same 
areas as those marked by the rectangles in Figs. 5a and 4b, respectively.}
\label{fig6}
\end{figure}

% \begin{table}
% \caption{}
% \label{}
% \end{tabular}
% \end{table}

\end{document}